\documentclass{svproc}

\usepackage{graphicx}
\usepackage[hidelinks]{hyperref}
\usepackage{multirow}
\usepackage{amsmath,amssymb,amsfonts}
\usepackage{mathrsfs}
\usepackage[title]{appendix}
\usepackage{xcolor}
\usepackage{textcomp}
\usepackage{manyfoot}
\usepackage{booktabs}
\usepackage{algorithm}
\usepackage{algorithmicx}
\usepackage{algpseudocode}
\usepackage{changepage}
\usepackage{listings}
\usepackage[utf8]{inputenc}
\usepackage{fourier}
\usepackage{array}
\usepackage{makecell}
\usepackage{xcolor}
\usepackage{tikz}
\usetikzlibrary{positioning}
\usepackage[misc]{ifsym}

\def\orcid#1{\kern .06em\href{https://orcid.org/#1}{\includegraphics[keepaspectratio,width=0.92em]{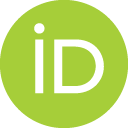}}}

\usepackage{url}

\newcommand{\vect}[1]{\boldsymbol{#1}}
\newcommand\Tstrut{\rule{0pt}{2.6ex}}         
\newcommand\Bstrut{\rule[-0.9ex]{0pt}{0pt}}   

\begin{document}

\mainmatter
\title{Influence Robustness of Nodes in Multiplex \\ Networks against Attacks}
\titlerunning{Influence Robustness of Nodes in Multiplex Networks against Attacks}

\author{
    Boqian Ma \orcid{0009-0007-5247-3001} \and
    Hao Ren \orcid{0000-0003-2169-0111} \and
    Jiaojiao Jiang \textsuperscript{\Letter} \orcid{0000-0001-7307-8114}
}
\authorrunning{Ma et al.}

\institute{School of Computer Science and Engineering, University of New South Wales, \\
           Kensington, NSW 2052, Australia\\
	\email{\{boqian.ma, hao.ren\}@student.unsw.edu.au}\\
	\email{jiaojiao.jiang@unsw.edu.au}
}
\maketitle

\begin{abstract}
    Recent advances have focused mainly on the resilience of the monoplex network in attacks targeting random nodes or links, as well as the robustness of the network against cascading attacks. However, very little research has been done to investigate the robustness of nodes in multiplex networks against targeted attacks. In this paper, we first propose a new measure, \texttt{MultiCoreRank}, to calculate the global influence of nodes in a multiplex network. The measure models the influence propagation on the core lattice of a multiplex network after the core decomposition. Then, to study how the structural features can affect the influence robustness of nodes, we compare the dynamics of node influence on three types of multiplex networks: assortative, neutral, and disassortative, where the assortativity is measured by the correlation coefficient of the degrees of nodes across different layers. We found that assortative networks have higher resilience against attack than neutral and disassortative networks. The structure of disassortative networks tends to break down quicker under attack.
\end{abstract}

\section{Introduction}\label{sec:introduction}

Many studies have been conducted to analyse the resilience of different types of networks, such as monoplex, interconnected, or multiplex networks, against different types of attacks (such as random, targeted, or cascading attacks). In monoplex networks, Albert \textit{et al.} \cite{albert2000error} found that networks with a broad degree distribution (such as scale-free) exhibit a low degree of resilience if the attack happened on a large degree node, and a high degree of resilience otherwise. A similar phenomenon occurs when cascading attacks occur in monoplex networks \cite{callaway2000network} \cite{motter2002cascade}. In interconnected networks, the malfunction of nodes within one network might trigger the collapse of reliant nodes in separate networks. Contrary to the behavior observed in single-layer networks, Buldyrev \textit{et al.} \cite{buldyrev2010catastrophic} demonstrated that a more heterogeneous degree distribution amplifies the susceptibility of independent networks to stochastic failures. Within the context of multiplex networks, which are composed of multiple layers sharing a common set of nodes \cite{bianconi2018multilayer}, various studies have indicated that correlated interconnections can influence the structural resilience of these networks in a complex manner \cite{min2014network} \cite{brummitt2015cascades}.

The resilience of a network against attacks is often measured from the perspective of network functionality, such as the probability of the existence of giant connected components \cite{buldyrev2010catastrophic}. However, this could not reflect the robustness of the node influence, which is of great significance. For example, in a power grid network, if a high-degree node is removed, many nodes that are connected to it will also be removed. Such removal will result in changes in the influence of the neighbouring nodes and beyond. To maintain the communication efficiency of a network, we often need to retain the robustness of influential nodes. Little research has been done to study the influence robustness of nodes in multiplex networks against attacks. In monoplex networks, Jiang \textit{et al.}\cite{jiang7841605} used the notion of coreness to measure the global influence of nodes. They found that nodes with high \textit{coreness} in assortative networks tend to maintain their degree of coreness even after the influential nodes are removed. On the other hand, in disassortative networks, the node's influence is distorted when influential nodes are removed.

In this paper, we extend the study of influence robustness of nodes against attacks from monoplex to multiplex networks. We first develop a new node centrality, \texttt{MultiCoreRank}, that measures the global influence of nodes in a multiplex network. Current centrality measures in multiplex networks are based on (1) projecting all layers into a monoplex network before applying the metrics on monoplex networks or (2) calculating the metrics in each individual layer separately, before aggregating to form a value for each node \cite{salehi2015spreading} \cite{de2015ranking}. However, these methods overlooked the multi-relation nature of multiplex networks, which could cause information loss in the process. To address this gap, we extend the idea of core decomposition in multiplex networks presented in \cite{galimberti2020core} and calculate the global influence of nodes through propagation of node influence along the ``core lattice''.

The main contributions of this paper include:
\begin{itemize}
	\item We propose, \texttt{MultiCoreRank}, a new node centrality that measures the global influence of nodes in multiplex networks.
	\item We analyse the influence robustness of nodes across different types of multiplex network: assortative, neutral, and disassortative networks. 
	\item The experimental results demonstrate that the assortative multiplex networks have greater robustness and are more resilient against targeted attack.
\end{itemize}

The rest of the paper is organised as follows. Section \ref{sec:related_work} introduces some related work. In Section \ref{sec:method_centrality}, we introduce the proposed centrality measure. Section \ref{sec:experiments} outlines our experimental results, followed by the conclusion in Section \ref{sec:conclusion}. In addition, code is available at {\color{blue}\url{https://github.com/Boqian-Ma/MultiCoreRank}}.

\section{Related Work}\label{sec:related_work}

Let $G=(V,E,L)$ be a multiplex network, where $V$ is a set of vertices, $L$ is a set of layers and $E\subseteq V \times V \times L$ is a set of links. Each layer of $G$ is a monoplex network $G^{[\alpha]}$, $\alpha\in L$. Each layer $\alpha$ is associated with an adjacency matrix $A^{[\alpha]}=\left(a^{[\alpha]}_{ij}\right)$, where $a^{[\alpha]}_{ij} = 1$ if there is a link between $i$ and $j$ on layer $\alpha$, and 0 otherwise. In the following, we first introduce some existing centrality measures, and then we discuss some related work on network resilience.

\begin{table}[b]
    \small
    \centering
    \caption{Node centralities in monoplex networks and multiplex networks. In the formula, $\lambda^{[\alpha]}$ represents the principal eigenvalue corresponding to the adjacency matrix $A^{[\alpha]}$. $\sigma_{pq}(i)$ signifies the aggregate number of shortest paths from node $p$ to node $q$ that traverse through node $i$, while $\sigma_{pq}$ indicates the overall number of shortest paths between nodes $p$ and $q$. $\mbox{dist}(i^{[\alpha]},j^{[\alpha]})$ is used to describe the minimal path distance between nodes $i$ and $j$ within layer $\alpha$.}
    \label{tab:notations}
    \begin{tabular}{l|c|c}
        \hline
        \textbf{Centrality} & \textbf{Monoplex}                                                                                                                             & \textbf{Multiplex} \Tstrut\Bstrut                          \\
        \hline
        Degree              & $d_i^{[\alpha]}=\sum_{j\in V} a_{ij}^{[\alpha]}$                                                                                              & $d_i = \sum_{\alpha\in L}d^{[\alpha]}_i$ \Tstrut\Bstrut    \\
        \hline
        Eigenvector         & $e_i^{[\alpha]} = \frac{1}{\lambda^{[\alpha]}} \sum_{j\in V} a^{[\alpha]}_{i,j} \, e^{[\alpha]}_{j}$                                          & $e_{i} = \sum_{\alpha\in L} e_i^{[\alpha]}$ \Tstrut\Bstrut \\
        \hline
        Betweenness         & $b_{i}^{[\alpha]} = \sum_{i \neq p \neq q \in V} \frac{\sigma_{p^{[\alpha]}q^{[\alpha]}}{(i^{[\alpha]})}}{\sigma_{p^{[\alpha]}q^{[\alpha]}}}$ & $b_i = \sum_{\alpha\in L} b_i^{[\alpha]}$ \Tstrut\Bstrut   \\
        \hline
        Closeness           & $c_i^{[\alpha]} = \frac{n-1}{\sum_{j\in V} \mbox{dist}(i^{[\alpha]},j^{[\alpha]})}$                                                           & $c_i = \sum_{\alpha\in L} c_i^{[\alpha]} $ \Tstrut\Bstrut  \\
        \hline
    \end{tabular}
\end{table}

\subsection{Node Centrality Measures}

Various centrality measures have been developed to calculate the influence of nodes on monoplex and multiplex networks. \textbf{Degree Centrality} quantifies the number of edges attached to a specific node in a monoplex network. It was was extended into \textbf{Overlapping Degree} in multiplex networks by summing the node's degree across various layers \cite{Batt14}. A node is considered influential if it is connected to a high number of edges. Bonacich \textit{et al.} formulated \textbf{Eigenvector Centrality}, and proposed that the principal eigenvector of an adjacency matrix serves as an effective indicator of a node's centrality within the network \cite{bonacich1971factoring}. Extending this to multiplex networks, Sola \textit{et al.} \cite{sola2013eigenvector} introduced multiple alternative metrics to evaluate the significance of nodes. \textbf{Betweenness centrality} measures the importance of a node by considering how often a node $v$ lies in a shortest path between $i$ and $j$ \cite{brandes2001faster}. Chakraborty \textit{et al.} \cite{chakraborty2016cross} extend betweenness centrality to multiplex networks and introduced cross-layer betweenness centrality. \textbf{Closeness Centrality} \cite{nieminen1974centrality} quantifies the proximity of a given node to all other nodes within a network by calculating the average distance via the shortest pathways to all other nodes. A node gains significant importance if it is situated closer to every other node within the network. Mittal \textit{et al.} \cite{mittal2018cross} introduced cross-layer closeness centrality for multiplex networks. We note the above measures as classical centrality measures and their counterparts on multiplex networks. 

More recently, other novel centrality methods have been proposed based on random walks \cite{chang2021boss}\cite{de2019general}, gravity model~\cite{curado2023novel}, and \textit{posteriori} measures \cite{lou2023structural}.


Table \ref{tab:notations} provides a list of classical centrality measures in monoplex and multiplex networks. Note that the counterpart of each centrality measure on a multiplex network is simply the sum of node centralities obtained on the different layers. For more complicated centrality measures, the readers can refer to \cite{de2013centrality}.

\subsection{Network Resilience}

Network resilience is measured by the ability of a network to retain its structure when some nodes in the network are removed \cite{cohen2010complex}. It can be measured by network assortativity, which describes the tendency of nodes in a network to connect to other nodes that are similar (or different) in some way. In recent decades, extensive contributions have been made to network resilience analysis \cite{callaway2000network,albert2000error,jiang7841605,motter2002cascade}. Understanding network resilience is of high research interest because it will allow us to design fail-safe networks such as transportation or energy networks.

In terms of the robustness of multilayer networks, Buldyrev \textit{et al.} \cite{buldyrev2010catastrophic} found that an interconnected network is vulnerable to random failures if it presents a broader degree distribution, which is the opposite of the phenomenon in monoplex networks. De \textit{et al.} \cite{de2014navigability} employed random walks to establish an analytical model for examining the time required for random walks to cover interconnected networks. Their findings indicate that such interconnected structures exhibit greater resilience to stochastic failures compared to their standalone layers. Min \textit{et al.} \cite{min2014network} studied the resilience of multiplex networks and found that correlated coupling can affect the structural robustness of multiplex networks in diversed fashion.  Brummitt \textit{et al.} \cite{brummitt2012multiplexity} generalised the threshold cascade model \cite{watts2002simple} to study the impact of multiplex networks on cascade dynamics. They found that multiplex networks are more vulnerable to global cascades than monoplex networks.

More recently, Fan \textit{et al.} \cite{fan2022modified} proposed a multiplex network resilience metric and studied link addition strategies to to improve resilience against targeted attacks. \\Kazawa \textit{et al.}~\cite{kazawa2020effectiveness} proposed effective link-addition strategies for improving the robustness of multiplex networks against degree-based attacks.

Recent studies mainly analyse resilience from a network functionality perspective, such as the probability of the existence of giant connected components \cite{buldyrev2010catastrophic}. This work extends from Jiang \textit{et al.}'s previous work on the influence robustness of nodes on monoplex networks. In this paper, we study the resilience of nodes in \textbf{multiplex networks} under \textbf{targeted} (i.e. attacking nodes based on their influence) and \textbf{uniformly random} attacks. Before that, we first develop a method to measure node influence based on core decomposition in multiplex networks (see Section \ref{sec:method_centrality}).

\section{Proposed Node Centrality}\label{sec:method_centrality}

\subsection{Preliminaries}\label{preliminaries}

Given a multiplex network $G=(V,E,L)$ and a subset $S \subseteq V$, we use $G[S] = (S, E[S], L)$ to denote the subgraph of $G$, where $E[S]$ is the set of all the links in $E$ connecting the nodes in $S$ and $L$ is the set of layers. We use $\tau^{[\alpha]}[S]$ to denote the minimum degree of nodes on layer $\alpha$ in the sub-graph
The core decomposition in multiplex networks is defined as follows.

\begin{definition}[$\boldsymbol{k}$-core percolation \cite{azimi2014k}] 
    Given a multiplex network $G=(V, E, L)$ and an $\vert L \vert$-dimensional integer vector $\boldsymbol{k} = [k^{[\alpha]}]_{\alpha \in L}$, the $\boldsymbol{k}$-core of $G$ is defined as the maximum subgraph $G[C] = (C, E[C], L)$ such that $\forall \alpha \in L: \tau^{[\alpha]}[C] \geq k^{[\alpha]}$. $\boldsymbol{k}$ is termed as a core vector.
\end{definition}

Hence, $\boldsymbol{k}$-core is the maximal subgraph that each node has at least $k^{[\alpha]}$ edges of each layer, $\alpha\in L$. The $\boldsymbol{k}$-core of a multiplex network could be calculated by removing nodes iteratively until $k^{[\alpha]} \in \boldsymbol{k}$, $\alpha\in L$ no longer satisfied. Taking the two-layer graph in Figure \ref{fig:example_network}(a) as an example, the $(1,2)$-core is $\{A, B, D ,E\}$ and the $(2,2)$-core is $\{B, D, E\}$.

\begin{theorem}[Core containment \cite{galimberti2020core}]\label{theorem:core_containment}
    Given a multiplex network $G=(V, E, L)$, let $C_{\boldsymbol{k}}$ and $C_{\boldsymbol{k'}}$ be the cores given by $\boldsymbol{k}=[k^{[\alpha]}]_{\alpha\in L}$ and $\boldsymbol{k'}=[k'^{[\alpha]}]_{\alpha\in L}$, respectively. It follows that if $\forall \alpha \in L: k'^{[\alpha]} \leq k^{[\alpha]}$, then $C_{\boldsymbol{k}} \subseteq C_{\boldsymbol{k'}}$.
\end{theorem}

The partial containment of all cores can be represented by a lattice structure known as the \textit{core lattice}. The core lattice of the example network in Figure \ref{fig:example_network}(a) is shown in Figure \ref{fig:core_lattice}. The nodes in the lattice represent cores and edges represent the containment relationship between cores where the ``father'' core contains all of its ``child'' cores (i.e. all cores from a core to the root). Using the core lattice structure, Galimberti \textit{et al.} \cite{galimberti2020core} developed three algorithms for efficiently computing cores in multiplex networks: DFS-based, BFS-based, and hybrid approaches. In the centrality method we are proposing in the next subsection, we use the BFS-based approach because, in order to update a node's influence given a core, all father cores must be calculated first. The approach based on Breadth-First Search (BFS) leverages two key observations: (1) a non-empty \(\boldsymbol{k}\)-core is a subset of the intersection of all its preceding cores "fathers", and (2) the quantity of such preceding cores for any non-empty \(\boldsymbol{k}\)-core is commensurate with the number of non-zero components in its associated core vector \(\boldsymbol{k}\).

\begin{figure}[!htb]
    \centering
    \begin{minipage}{0.45\textwidth}
        \centering
        \begin{tikzpicture}[node distance={20mm}, thick, main/.style = {draw, circle}]
            \node[main] (1) {$A$}; 
            \node[main] (2) [right of=1] {$B$};
            \node[main] (3) [right of=2] {$C$}; 
            \node[main] (4) [below of=1] {$D$};
            \node[main] (5) [right of=4] {$E$}; 
            \node[main] (6) [right of=5] {$F$};
            
            \draw (1) -- (2);
            \draw (2) -- (3);
            \draw (2) -- (4);
            \draw (2) -- (5);
            \draw (2) -- (6);
            \draw (3) -- (5);
            \draw (3) -- (6);
            \draw (3) -- (6);
            \draw (4) -- (5);
            \draw (5) -- (6);
            
            \draw[dashed] (1) to [out=20,in=160,looseness=0.5] (2);
            \draw[dashed] (1) to [out=250,in=110,looseness=0.5] (4);
            \draw[dashed] (2) to [out=200,in=65,looseness=0.5] (4);
            \draw[dashed] (2) to [out=250,in=110,looseness=0.5] (5);
            \draw[dashed] (3) to [out=200,in=65,looseness=0.5] (5);
            \draw[dashed] (4) to [out=340,in=200,looseness=0.5] (5);
            \draw[dashed] (5) to [out=340,in=200,looseness=0.5] (6);
            \end{tikzpicture} 
        \caption{An example two-layer network, where solid lines signify edges belonging to the first layer, while dashed lines indicate edges associated with the second layer.}
        \label{fig:example_network}
    \end{minipage}\hfill
    \begin{minipage}{0.45\textwidth}
        \centering
        \begin{tikzpicture}[node distance={20mm}, thick, main/.style = {draw, circle}]
        \node[main] (1) {$A$}; 
        \node[main] (2) [right of=1] {$B$};
        \node[main] (3) [right of=2] {$C$}; 
        \node[main] (4) [below of=1] {$D$};
        \node[main] (5) [right of=4] {$E$}; 
        \node[main] (6) [right of=5] {$F$};
        
        \draw (3) -- (5);
        \draw (3) -- (6);
        \draw (3) -- (6);
        \draw (4) -- (5);
        \draw (5) -- (6);
        
        \draw[dashed] (1) to [out=250,in=110,looseness=0.5] (4);
        \draw[dashed] (3) to [out=200,in=65,looseness=0.5] (5);
        \draw[dashed] (4) to [out=340,in=200,looseness=0.5] (5);
        \draw[dashed] (5) to [out=340,in=200,looseness=0.5] (6);

        \end{tikzpicture}
        \caption{The network after removing node $B$ and its edges in Figure \ref{fig:example_network}.}
        \label{fig:after_removal}
    \end{minipage}
\end{figure}

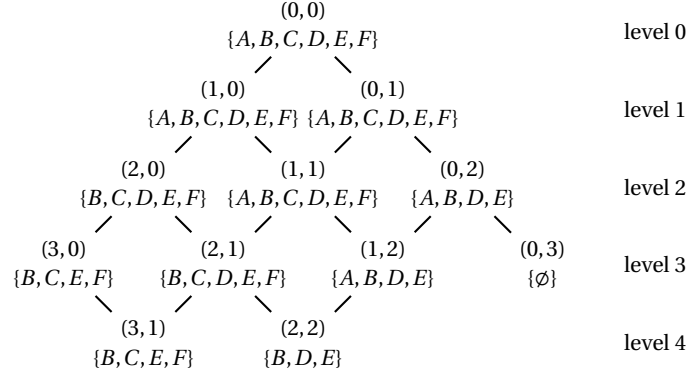
\begin{figure}[!htb]
    \centering
    \begin{tikzpicture}[node distance={15mm}, thick, main/.style = {draw, circle}]
        \node[draw=none,fill=none, align=center] (1) {$(0,0)$ \\ $\{A,B, C, D, E, F\}$};
        
        \node[draw=none,fill=none, align=center] (2) [below left of=1]{$(1,0)$ \\ $\{A, B, C, D, E, F\}$};
        \node[draw=none,fill=none, align=center] (3) [below right of=1]{$(0,1)$ \\ $\{A, B, C, D, E, F\}$};
        
        
        \node[draw=none,fill=none, align=center] (4) [below left of=2]{$(2,0)$ \\ $\{B, C, D, E, F\}$};
        \node[draw=none,fill=none, align=center] (5) [below right of=2]{$(1,1)$ \\ $\{A, B, C, D, E, F\}$};
        \node[draw=none,fill=none, align=center] (6) [below right of=3]{$(0,2)$ \\ $\{A, B, D, E\}$};
        
        \node[draw=none,fill=none, align=center] (7) [below left of=4]{$(3,0)$ \\ $\{B, C, E, F\}$};
        \node[draw=none,fill=none, align=center] (8) [below right of=4]{$(2,1)$ \\ $\{B, C, D, E, F\}$};
        \node[draw=none,fill=none, align=center] (9) [below left of=6]{$(1,2)$ \\ $\{A, B, D, E\}$};
        \node[draw=none,fill=none, align=center] (11) [below right of=6]{$(0,3)$ \\ $\{\emptyset\}$};

        \node[draw=none,fill=none, align=center] (10) [below right of=8]{$(2,2)$ \\ $\{B, D, E\}$};
        
        \node[draw=none,fill=none, align=center] (12) [below right of=7]{$(3,1)$ \\ $\{B, C, E, F\}$};
        
        
        \node[draw=none,fill=none, align=center] (17) [right of=11]{level 3};
        =
        right=3cm of B
        \node[draw=none,fill=none, align=center] (18) [below=6mm of 17]{level 4};
        \node[draw=none,fill=none, align=center] (16) [above=6mm of 17]{level 2};
        \node[draw=none,fill=none, align=center] (15) [above=6mm of 16]{level 1};
        \node[draw=none,fill=none, align=center] (14) [above=6mm of 15]{level 0};

        \draw (1) -- (2);
        \draw (1) -- (3);
        \draw (2) -- (4);
        \draw (2) -- (5);
        \draw (3) -- (5);
        \draw (3) -- (6);
        \draw (4) -- (7);
        \draw (4) -- (8);
        \draw (5) -- (8);
        \draw (5) -- (9);
        \draw (6) -- (9);
        \draw (6) -- (11);
        \draw (8) -- (10);
        \draw (9) -- (10);
        
        \draw (8) -- (12);
        \draw (7) -- (12);
        
        \end{tikzpicture}

    \caption{The core lattice of the network in Figure \ref{fig:example_network}. The numbers in the core vectors are the minimum degrees of each layer in a core. (i.e. (2,1) consists of nodes with at least degree 2 on layer 1 and at least degree 1 on layer 2.)
    The level in which a core is in the lattice is given by the \textit{L-1} norm of its core vector. The core vector (0,3) is shown as an example of empty core while other empty cores are omitted.}
    \label{fig:core_lattice}
\end{figure}

\subsection{MultiCoreRank Centrality} \label{proposed method}

On a core lattice, we can observe that (1) nodes that appear in deeper level cores are more connected, which makes them more influential than those that only appear in shallower levels, and (2) for nodes on the same lattice level, those that appear in fewer cores are less influential than those that appear in more cores, as they have higher chances to have child cores, according to the core containment theory in Theorem \ref{theorem:core_containment}. We argue that an \textit{ideal} centrality measure based on the core lattice should at least consider these two points.

Using the two observations above, we consider the calculation of the overall influence of a node $v\in V$ as a message passing process, which iteratively calculates the influence of node $v$ on layer $l+1$ based on its influence on level $l$ on the lattice.

Before introducing \texttt{MultiCoreRank}, we first define a father-child relationship, $\vect{k}_{l+1} \xrightarrow[\text{}]{\text{father}} \vect{k}_{l}$, of the two corresponding core vectors on levels $l+1$ and $l$, respectively, if there exists an edge between $\vect{k}_{l+1}$ and $\vect{k}_{l}$ on the core lattice. For an arbitrary node $v\in V$, we use $\mbox{inf}_{l}(v)$ to denote the influence of node $v$ on level $l$ of the core lattice, if there exists a $\vect{k}_l$-core such that $v\in C_{\vect{k}_l}$. Also, let $|\mbox{father}_{\vect{k}_{l}}|$ be the number of fathers of the core given by $\vect{k}_{l}$.

Now, considering observation (1), for an arbitrary level $l$, we assign it a weight $l$, allowing nodes at deeper lattice levels to have a larger weight.
Next, considering observation (2), for an arbitrary node $v$, if $v$ appears at both level $l$ and level $l+1$ on the lattice, we aggregate $v$'s influences from all the cores that contain node $v$ on level $l$ as its influence on level $l+1$.
In this way, nodes with more appearances will be assigned a higher influence than those with fewer appearances.
The following formula gives the detailed calculation of the \texttt{MultiCoreRank} influence of a node $v$ on level $l+1$ of a core lattice:
\begin{equation}\label{eq.inf_v}
    \begin{array}{ll}
        \mbox{inf}_{l+1}(v) = \sum_{\substack{\big\{\vect{k}_{l+1} \big{\vert} v\in C_{\vect{k}_{l+1}}, v\in C_{\vect{k}_{l}},  \vect{k}_{l+1} \xrightarrow[\text{}]{\text{father}} \vect{k}_{l} \big\}}} (l+1) \cdot \mbox{inf}_{l}(v) \cdot \mbox{inf}(C_{\vect{k}_{l}}) \cdot \lvert \mbox{father}_{\vect{k}_{l+1}}\rvert,
    \end{array}
\end{equation}
where $\mbox{inf}(C_{\vect{k}_{l}})$ is the influence of the $\vect{k}_{l}$-core, calculated by
\begin{equation}
    \mbox{inf}(C_{\vect{k}_{l}}) = \frac{\sum_{v\in C_{\vect{k}_l}}\mbox{inf}_l(v)}{\lvert C_{\vect{k}_l} \rvert}.
\end{equation}

Note that our proposed influence measure can be calculated using the BFS-based approach mentioned in Section \ref{preliminaries} because of the layer-by-layer and iterative nature of this method. The influence of a node $v$ on layer $l+1$ is calculated on the basis of its influence on level $l$, hence all cores on layer $l$ must be calculated before moving onto layer $l+1$. Referring to Figure \ref{fig:core_lattice}, the order in which the cores are calculated from layer 0 to 2 is $(0,0), (1,0), (0,1), (2,0), (1,1), (0,2)$.

To illustrate our method, consider node $A$ in Figure \ref{fig:example_network} and the lattice in Figure \ref{fig:core_lattice}. At level 0 of the lattice, since the $(0,0)$-core contains the entire network, we initialise the influence of all nodes to 1. On level 1, after the $(1,0)$-core and the $(0,1)$-core are found using BFS, we apply Eq. (\ref{eq.inf_v}) to node $A$ to get
\begin{align*}
	\mbox{inf}_{l=1}(A) & =  \sum_{\vect{k}_{l} \in \{(0,1), (1,0)\}} 1 \cdot \mbox{inf}_{l=0}(A) \cdot \mbox{inf}(C_{\vect{k}_{l-1}}) ~~~ \cdot \lvert \mbox{fathers}_{\vect{k}_{l}} \rvert =  1 + 1 = 2.
\end{align*}

On level 2, we have the following equation:
\begin{align*}
    \mbox{inf}_{l=2}(A) & = \sum_{\vect{k}_{l} \in \{(1,1), (0,2)\}} 2 \cdot \mbox{inf}_{l=1}(A) \cdot \mbox{inf}(C_{(\vect{k}_{l-1}}) \cdot \lvert \mbox{fathers}_{\vect{k}_{l}} \rvert \\
                        & = (2 \cdot 2 \cdot 2 \cdot 2) + (2 \cdot 2 \cdot 2 \cdot 1) = 24.
\end{align*}
The rest can be deduced accordingly \footnote{When implementing this method on a large-scale network, appropriate normalisation techniques are required when the network is large to prevent numeric overflow.}.

\section{Empirical Analysis of Influence Robustness of Nodes in Multiplex Networks}\label{sec:experiments}

In this section, we commence by delineating the assortativity metric employed for gauging the structural characteristics of multiplex networks. Subsequently, we provide an overview of the data sets utilized for experimental validation. Following that, we assess the performance efficacy of the proposed centrality metric for nodes. Lastly, we undertake an analysis of the robustness of multiplex networks under varying levels of assortativity.

\subsection{Multiplex Network Assortativity}

We study the resilience of multiplex networks by analysing the dynamics of node influence when the most influential nodes are removed. We particularly consider the structural feature of \textit{assortativity} of multiplex networks.

The assortativity of a multiplex network is measured by the average layer-layer degree correlation \cite{nicosia2015measuring}. If we denote $\vect{d}^{[\alpha]} = (d_{1}^{[\alpha]}, \cdots, d_{|V|}^{[|L|]})$ and $\vect{d}^{[\beta]} = (d_{1}^{[\beta]}, \cdots, d_{|V|}^{[|L|]})$ as the degree vectors of layer $\alpha$ and $\beta$ respectively, the layer-layer degree correlation between these two layers is given by
\begin{equation}\label{layer-layer}
    r_{\alpha, \beta} = \frac{\langle \boldsymbol{d}^{[\alpha]} \boldsymbol{d}^{[\beta]}\rangle - \langle \boldsymbol{d}^{[\alpha]}\rangle \langle \boldsymbol{d}^{[\beta]}\rangle}{\sigma_{\boldsymbol{d}^{[\alpha]}} \sigma_{\boldsymbol{d}^{[\beta]}}},
\end{equation}
where $\sigma_{\boldsymbol{d}^{[\alpha]}} = \sqrt{\langle\boldsymbol{d}^{[\alpha]} \boldsymbol{d}^{[\alpha]}\rangle - {\langle \boldsymbol{d}^{[\alpha]} \rangle}^2}$.  $r_{\alpha, \beta}$ is the Spearman coefficient of $\vect{d}^{[\alpha]}$ and $\vect{d}^{[\beta]}$. $r_{\alpha, \beta}$ being close to $1$ (assortative) means that the nodes in layers $\alpha$ and $\beta$ are likely to have a similar tendency when connecting with their neighbours (i.e. a node has relative high/low degrees in both layers), whereas $r_{\alpha, \beta}$ being close to $-1$, means that the nodes in $\alpha$ and $\beta$ are less likely to have a similar tendency when connecting with their neighbours (i.e. a node has relative high/low degrees in one layer by the opposite in the other layer).

In this paper, we compute the assortativity of a multiplex network as the average of all layer-layer degree correlation given by
\begin{equation}\label{eq.r_G}
    r_{G} = \frac{\sum_{\alpha < \beta \in L} r_{\alpha,\beta} }{|L|^2 - L}
\end{equation}
Without losing generality, we disregards all correlations of a layer to itself.

\subsection{Datasets}

We selected two datasets for each of the assortative, neutural, and disassortative networks for our experiments. The details of the datasets are as follows, and Table \ref{tab:statistics} gives the basic statistics of the datasets. \textbf{C.elegans}\footnote{\url{https://manliodedomenico.com/data.php}} \cite{nicosia2015measuring} is the neural network of the C.elegans nematode worm that consists of two layers representing synapses and gap junctions.
\textbf{Aarhus} \cite{magnani2013combinatorial} is a five-layer network that encapsulates five different types of interactions (Facebook, Leisure, Work, Co-authorship, and Lunch) among employees within the Computer Science department at Aarhus University. \textbf{OpenFlight continental airport networks}\footnote{\url{https://openflights.org/}} \cite{nicosia2015measuring} consists of international flight routes, where layers represent an airline company, node represent airports and edges represent routes provided by the airlines.  We selected layers of \textbf{South America} and \textbf{North America} such that the network is disassortative and layers of \textbf{Asia} and \textbf{Europe} such that the network is neutral.

\begin{table}[h!]
    \small
    \centering
    \caption{Statistics of the datasets used in this paper, where $r_G$ is the assortativity calculated from Equation (\ref{eq.r_G}).}
    \label{tab:statistics}
    \begin{tabular}{l|c|c|c|c|c}
        \hline
        \textbf{Dataset}                & \textbf{Network} & \textbf{\#nodes} & \textbf{\#links} & \textbf{\# Selected layers} & $\boldsymbol{r_G}$ \Tstrut\Bstrut \\
        \hline
        C.elegans                       & Assortative      & 281              & 2476             & 2                           & 0.6414 \Tstrut\Bstrut             \\
        \hline
        Aarhus                          & Assortative      & 61               & 620              & 5                           & 0.2160 \Tstrut\Bstrut             \\
        \hline
        Airlines-Europe (Air-EU)        & Neutral          & 476              & 3068             & 75                          & 0.0139 \Tstrut\Bstrut             \\
        \hline
        Airlines-Asia (Air-Asia)        & Neutral          & 348              & 1281             & 63                          & 0.0125 \Tstrut\Bstrut             \\
        \hline
        Airlines-SouthAmerica (Air-SAM) & Disassortative   & 129              & 272              & 13                          & -0.0141 \Tstrut\Bstrut            \\
        \hline
        Airlines-NorthAmerica (Air-NAM) & Disassortative   & 528              & 1699             & 33                          & -0.0052 \Tstrut\Bstrut            \\
        \hline
    \end{tabular}
\end{table}

\begin{figure*}[!htb]
    \centering
    \includegraphics[width=\textwidth]{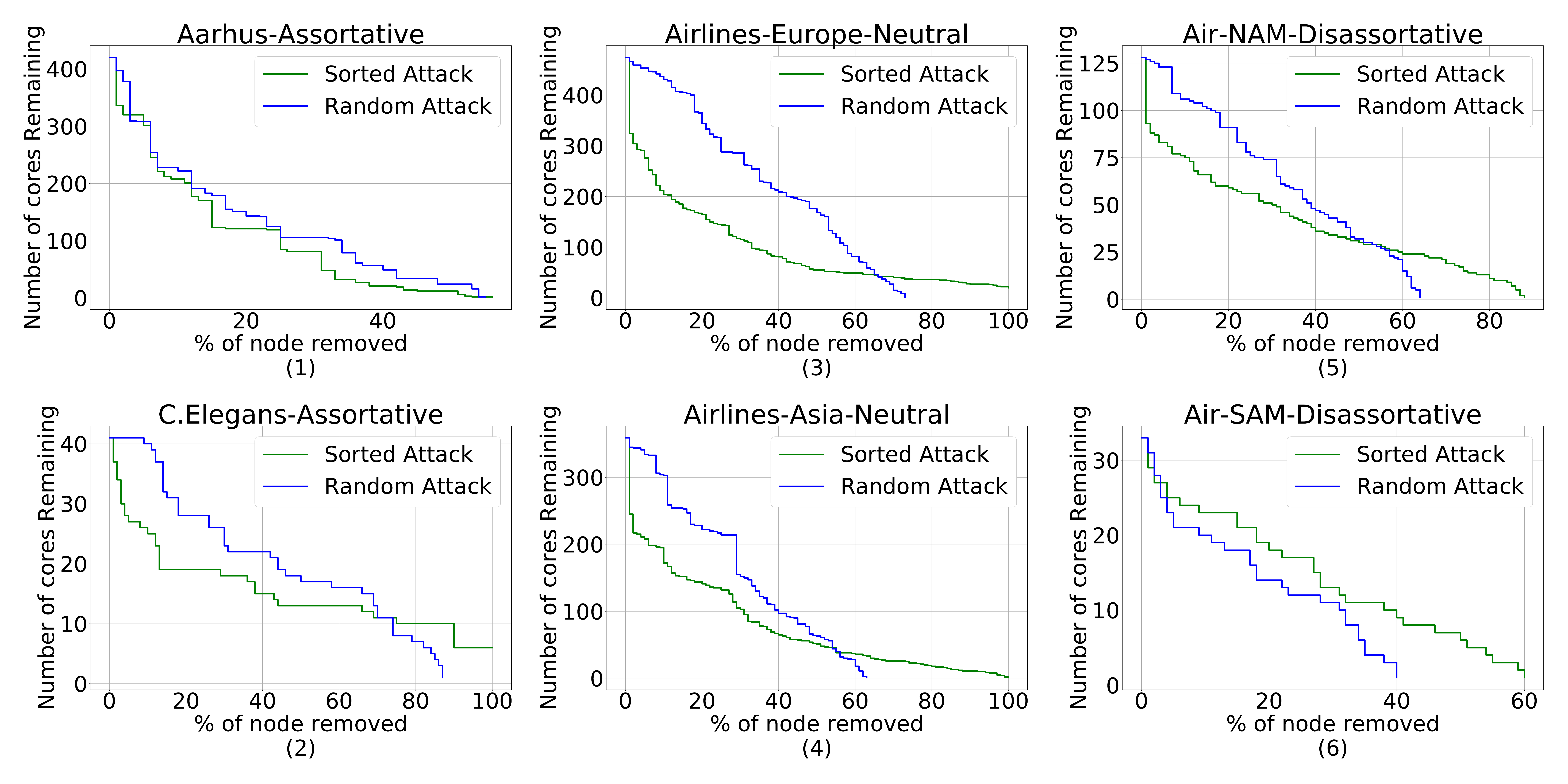}
    \caption{The number of cores remaining in the network after a percentage of nodes are removed. Two types of removal are performed 1) sorted attack based on \texttt{MultiCoreRank} and 2) uniformly random attack. (1) and (2) correspond to the two assortative networks, (3) and (4) correspond to the two neutral networks, and (5) and (6) correspond to the two disassortative networks.}
    \label{fig:number_of_cores}
\end{figure*}

\begin{figure*}[ht]
    \centering
    \includegraphics[width=\textwidth]{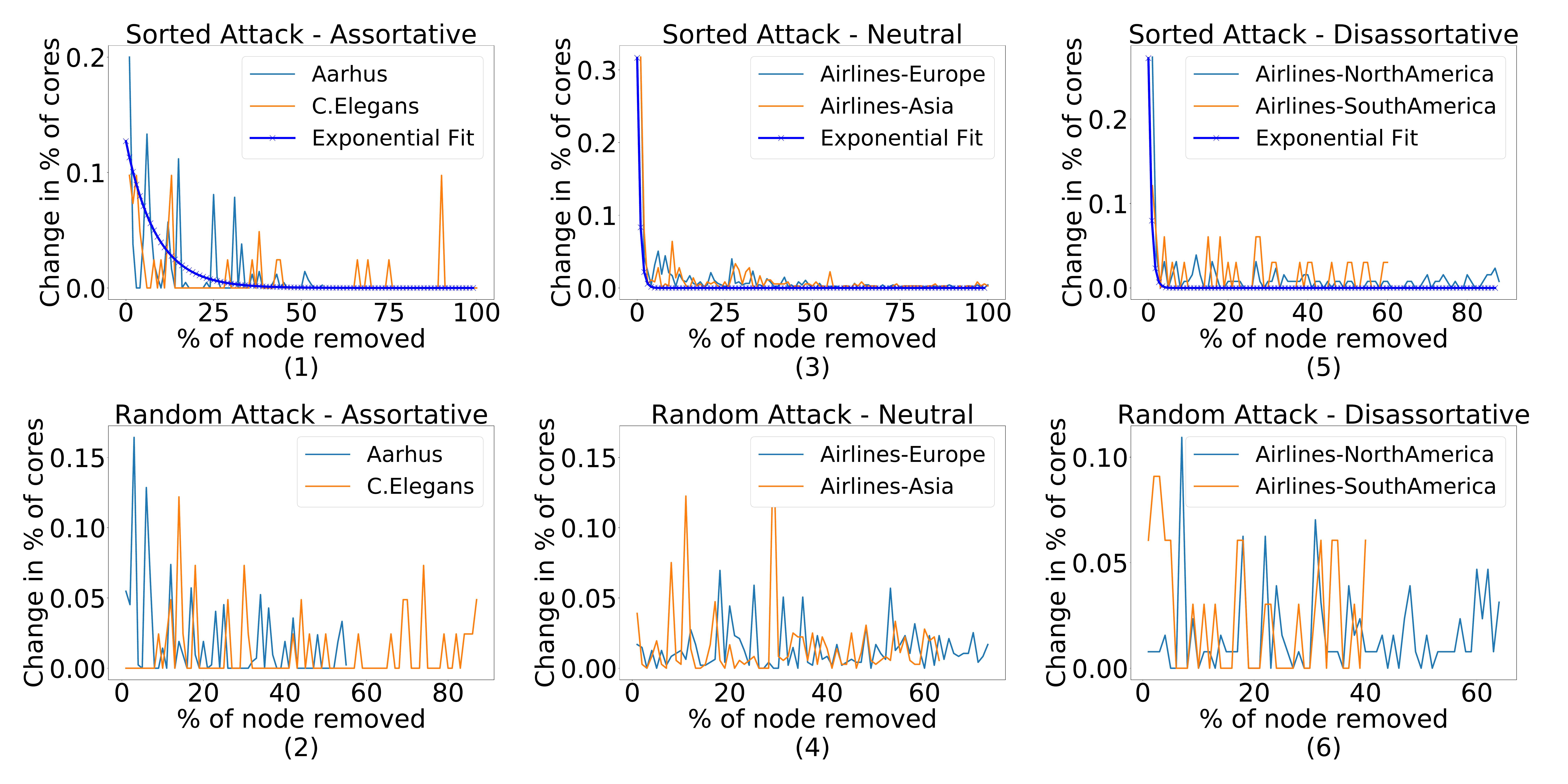}
    \caption{The change in percentage of cores as a percentage of nodes are removed in different types of networks. The top row shows the sorted attack results. The bottom row shows the random attack results. The exponential fit on the top row is an exponential function,$y=a\mathrm{e}^{-xb}$, fitted on the average y-value given by each dataset at each x-value.}
    \label{fig:slops}
\end{figure*}

\subsection{Effectiveness of the Proposed Centrality}

To evaluate the effectiveness of our method, we calculated the Spearman's Coefficient between our measurement and other node centralities. The results are shown in Table \ref{tab:spearman_effectiveness}. We found that our method correlates the most with overlapping degree ($d_i$). This is justifiable since $\vect{k}$-core is obtained by removing nodes that no longer satisfy the degrees given by a coreness vector, leaving only the nodes with higher degrees. Hence, our method is also based on the degree of a node. For the assortative and neutral datasets, our method has shown strong correlations with the eigenvector, betweenness, and closeness centralities. However, with the disassortative datasets, SouthAmerica and NorthAmerica, eigenvector and betweenness centrality show relatively weak correlation.

In addition, we compare the effectiveness of our method when random and influential nodes are removed from the network. Figure \ref{fig:number_of_cores} simulates the changes in the network structure when proportions of nodes is removed, which simulate attacks. In general, the quality of our centrality measure is demonstrated through the sharp decrease in the number of cores at the beginning of the attacks (i.e. Figure \ref{fig:number_of_cores} (3, 4, 5, 6)). This corresponds to the effectiveness of our method in identifying highly influential nodes because removing them caused significant structural changes.

\begin{table}[!tb]
    \small
    \centering
    \parbox{.45\linewidth}{
        \centering
        \caption{Spearman's rank correlation coefficients between node influence rankings from our method and other centrality measures in a complete network.}
        \label{tab:spearman_effectiveness}
        \begin{tabular}{l|c|c|c|c}
            \hline
            \textbf{Dataset} & $\boldsymbol{d_i}$ & $\boldsymbol{\lambda_i}$ & $\boldsymbol{b_i}$ & $\boldsymbol{c_i}$ \Tstrut\Bstrut \\
            \hline
            C.Elegans        & 0.85               & 0.65                     & 0.65               & 0.78 \Tstrut\Bstrut               \\
            \hline
            Aarhus           & 0.82               & 0.76                     & 0.49               & 0.81 \Tstrut\Bstrut               \\
            \hline
            Air-EU           & 0.51               & 0.45                     & 0.38               & 0.55 \Tstrut\Bstrut               \\
            \hline
            Air-Asia         & 0.76               & 0.56                     & 0.50               & 0.72 \Tstrut\Bstrut               \\
            \hline
            Air-SAM          & 0.93               & 0.29                     & 0.62               & 0.86 \Tstrut\Bstrut               \\
            \hline
            Air-NAM          & 0.83               & 0.37                     & 0.49               & 0.60 \Tstrut\Bstrut               \\
            \hline
        \end{tabular}
    }
    \hfill
    \parbox{.45\linewidth}{
        \centering
        \caption{Calculated coefficients of degree correlation between network layers following the targeted removal of the top 10\%, 20\%, and 30\% of nodes.}
        \label{tab:assortativity_10_20_30}
        \begin{tabular}{l|c|c|c|c}
            \hline
            \textbf{Dataset} & $\textbf{0\%}$ & $\textbf{10\%}$ & $\textbf{20\%}$ & $\textbf{30\%}$ \Tstrut\Bstrut \\
            \hline
            C.Elegans        & 0.6414         & 0.4174          & 0.4182          & 0.4315 \Tstrut\Bstrut          \\
            \hline
            Aarhus           & 0.2163         & 0.2723          & 0.3364          & 0.4110 \Tstrut\Bstrut          \\
            \hline
            Air-EU           & 0.0139         & 0.0061          & 0.0081          & 0.0042 \Tstrut\Bstrut          \\
            \hline
            Air-Asia         & 0.0125         & 0.0065          & 0.0072          & 0.0057 \Tstrut\Bstrut          \\
            \hline
            Air-SAM          & -0.0141        & -0.0050         & 0.0005          & -0.0273 \Tstrut\Bstrut         \\
            \hline
            Air-NAM          & -0.0052        & -0.0042         & -0.0057         & -0.0062 \Tstrut\Bstrut         \\
            \hline
        \end{tabular}
    }
\end{table}

\subsection{Influence Robustness of Nodes}

Continuing on the node removal experiment from the previous section, we also analyse the impact on the overall network assortativity when nodes are removed. Figure \ref{fig:slops} (1,3,5) shows the change in the percentage of nodes when the influential nodes are removed in order. Comparing Figure \ref{fig:slops} (1) with (3,5), we see that the change in the number of remaining cores is less drastic in the assortative networks when influential nodes are removed. That is to say, 1) assortative networks are more robust under targeted attacks, 2) the removal of high-influence nodes in a neutral and disassortative network has a high impact on the network robustness. On the other hand, when nodes are removed randomly as shown in Figure \ref{fig:slops} (2,4,6), the network structure does not show visible trends in terms of changes.

Table \ref{tab:assortativity_10_20_30} presents the change in assortativity when a percentage of high-influence nodes are removed. The assortative networks (C.Elegans, Aarhus) remained relatively assortative after node removal. The disassortative networks remained disassortative. The neutral networks remained neutral. However, there is a trend in decreasing in assortativity in all types of network. This suggests that the initial network assortativity is a good indicator of the robustness of a given network.

Overall, from the above experiments, we can see that, similar to monoplex networks, assortative networks have shown higher robustness against attack than neutral and disassortative networks. The change in the overall $\vect{k}$-core structure of the networks is smaller for the assortative networks.

\section{Conclusion}\label{sec:conclusion}

In summary we developed a new node centrality measure, \texttt{MultiCoreRank} node centrality, based on core decomposition in multiplex networks. This measure takes into account the multi-relation nature of such networks and has shown consistency with existing methods through empirical comparisons. We then analysed the influence robustness of nodes across different types of multiplex networks: assortative, neutral and disassortative networks. We found that, in assortative networks, the $\vect{k}$-core structure remains more consistent when nodes of high influence are removed. However, in neutral and disassortative networks, the number of $\vect{k}$-cores tends to quickly decrease when they are under attack. In future work, we aim to  study defence mechanisms to increase the robustness of multiplex networks and extend our method to multi-layer networks.

\bibliographystyle{spmpsci} 
\bibliography{refs} 

\end{document}